\documentstyle[12pt]{article}
\begin{document}
\title{Non-Noether symmetries and their influence on phase space geometry}
\author{George Chavchanidze} \date{} \maketitle
\thanks{Department of Theoretical Physics, A. Razmadze Institute of Mathematics, 1 Aleksidze Street, Tbilisi 0193, Georgia}
\begin{abstract}{\bf Abstract.} We disscuss some geometric aspects of the concept of non-Noether symmetry. It is shown that in regular Hamiltonian systems such a symmetry canonically leads to a Lax pair on the algebra of linear operators on cotangent bundle over the phase space. Correspondence between the non-Noether symmetries and other wide spread geometric methods of generating conservation laws such as bi-Hamiltonian formalism, bidifferential calculi and Fr\"{o}licher-Nijenhuis geometry is considered. It is proved that the integrals of motion associated with the continuous non-Noether symmetry are in involution whenever the generator of the symmetry satisfies a certain Yang-Baxter type equation.\end{abstract}
In the present paper we would like to shed more light on geometric aspects of the concept of non-Noether symmetry and to emphasize influence of such a symmetries on the phase space geometry. Partially the motivation for studying these issues comes from the theory of integrable models that essentially relies on different geometric objects used for constructing conservation laws. Among them are Fr\"{o}licher-Nijenhuis operators, bi-Hamiltonian systems, Lax pairs and bicomplexes. And it seems that the existance of these important geometric structures could be related to the hidden non-Noether symmetries of the dynamical systems. We would like to show how in Hamiltonian systems presence of certain non-Noether symmetries leads to the above mentioned Lax pairs, Fr\"{o}licher-Nijenhuis operators, bi-Hamiltonian structures, bicomplexes and a number of conservation laws. \\
Let us first recall some basic knowledge of the Hamiltonian dynamics. The phase space of a regular Hamiltonian system is a Poisson manifold -- a smooth finite-dimensional manifold equipped with the Poisson bivector field $W$ subjected to the following condition 
\begin{eqnarray} \label{eq:e1}
[W , W] = 0
\end{eqnarray}
where square bracket stands for Schouten bracket or supercommutator (for simplicity further it will be referred as commutator). In a standard manner Poisson bivector field defines a Lie bracket on the algebra of observables (smooth real-valued functions on phase space) called Poisson bracket: 
\begin{eqnarray}
\{f , g\} = W(df \wedge dg)
\end{eqnarray}
Skew symmetry of the bivector field $W$ provides the skew symmetry of the corresponding Poisson bracket and the condition (\ref{eq:e1}) ensures that for every triple $(f, g, h)$ of smooth functions on the phase space the Jacobi identity 
\begin{eqnarray}
\{f\{g , h\}\} + \{h\{f , g\}\} + \{g\{h , f\}\} = 0.
\end{eqnarray}
is satisfied. We also assume that the dynamical system under consideration is regular -- the bivector field $W$ has maximal rank, i. e. its $n$-th outer power, where $n$ is a half-dimension of the phase space, does not vanish $W^{n} \neq 0$. In this case $W$ gives rise to a well known isomorphism $\Phi _{W}$ between the differential 1-forms and the vector fields defined by 
\begin{eqnarray} \label{eq:e2}
\Phi _{W}(u) = W(u)
\end{eqnarray}
for every 1-form $u$ and could be extended to higher degree differential forms and multivector fields by linearity and multiplicativity $\Phi _{W}(u \wedge v) = \Phi _{W}(u) \wedge \Phi _{W}(v)$. \\
Time evolution of observables is governed by the Hamilton's equation 
\begin{eqnarray} \label{eq:e3}
\frac{d}{dt}f = \{h , f\}
\end{eqnarray}
where $h$ is some fixed smooth function on the phase space called Hamiltonian. Let us recall that each vector field $E$ on the phase space generates the one-parameter continuous group of transformations $g_{a} = e^{aL_{E}}$ (here $L$ denotes Lie derivative) that acts on the observables as follows 
\begin{eqnarray}
g_{a}(f) = e^{aL_{E}}(f) = f + aL_{E}f + \frac{1}{2}a^{2}L_{E}^{2}f + ...
\end{eqnarray}
Such a group of transformation is called symmetry of Hamilton's equation (\ref{eq:e3}) if it commutes with time evolution operator 
\begin{eqnarray}
\frac{d}{dt} g_{a}(f) = g_{a}(\frac{d}{dt}f) 
\end{eqnarray}
in terms of the vector fields this condition means that the generator $E$ of the group $g_{a}$ commutes with the vector field $W(h) = \{h , \}$, i. e. 
\begin{eqnarray} \label{eq:e4}
[E , W(h)] = 0.
\end{eqnarray}
However we would like to consider more general case where $E$ is time dependent vector field on phase space. In this case (\ref{eq:e4}) should be replaced with 
\begin{eqnarray} \label{eq:e5}
\frac{ \partial }{\partial t }E = [E , W(h)].
\end{eqnarray}
If in addition to (\ref{eq:e4}) the vector field $E$ does not preserve Poisson bivector field $[E , W] \neq 0$ then $g_{a}$ is called non-Noether symmetry.\\
Now let us focus on non-Noether symmetries. We would like to show that the presence of such a symmetry could essentially enrich the geometry of the phase space and under the certain conditions could ensure integrability of the dynamical system. Before we proceed let us recall that the non-Noether symmetry leads to a number of integrals of motion \cite{r4}. More precisely the relationship between non-Noether symmetries and the conservation laws is described by the following theorem. \\
{\bf Theorem 1.} Let $(M , h)$ be regular Hamiltonian system on the $2n$-dimensional Poisson manifold $M$. Then, if the vector field $E$ generates non-Noether symmetry, the functions 
\begin{eqnarray} \label{eq:e6}
Y^{(k)} = \frac{ \hat{W}^{k} \wedge W^{n - k} }{W^{n} } ~~~~~k = 1,2, ... n
\end{eqnarray}
where $\hat{W} = [E , W]$, are integrals of motion. \\
{\bf Proof:} By the definition 
\begin{eqnarray}
\hat{W}^{k} \wedge W^{n - k} = Y^{(k)}W^{n}.
\end{eqnarray}
(definition is correct since the space of $2n$ degree multivector fields on $2n$ degree manifold is one dimensional). Let us take time derivative of this expression along the vector field $W(h)$, 
\begin{eqnarray}
\frac{d}{dt}\hat{W}^{k} \wedge W^{n - k} = (\frac{d}{dt}Y^{(k)})W^{n} + Y^{(k)}[W(h) , W^{n}]
\end{eqnarray}
or 
\begin{eqnarray} \label{eq:e7}
k(\frac{d}{dt}\hat{W}) \wedge \hat{W}^{k - 1} \wedge W^{n - k} + (n - k)[W(h) , W] \wedge \hat{W}^{k} \wedge W^{n - k - 1} = \nonumber \\(\frac{d}{dt}Y^{(k)})W^{n} + nY^{(k)}[W(h) , W] \wedge W^{n - 1}
\end{eqnarray}
but according to the Liouville theorem the Hamiltonian vector field preserves $W$ i. e. 
\begin{eqnarray}
\frac{d}{dt}W = [W(h) , W] = 0
\end{eqnarray}
hence, by taking into account that 
\begin{eqnarray}
\frac{d}{dt}E= \frac{ \partial }{\partial t }E + [W(h) , E] = 0
\end{eqnarray}
we get 
\begin{eqnarray}
\frac{d}{dt}\hat{W} = \frac{d}{dt}[E , W] = [\frac{d}{dt}E, W] + [E[W(h) , W]] = 0.
\end{eqnarray}
and as a result (\ref{eq:e7}) yields 
\begin{eqnarray}
\frac{d}{dt}Y^{(k)}W^{n} = 0
\end{eqnarray}
but since the dynamical system is regular ($W^{n} \neq 0$) we obtain that the functions $Y^{(k)}$ are integrals of motion. \\
{\bf Remark.} Instead of conserved quantities $Y^{(1)} ... Y^{(n)}$, the solutions $c_{1} ... c_{n}$ of the secular equation 
\begin{eqnarray} \label{eq:e8}
(\hat{W} - cW)^{n} = 0
\end{eqnarray}
could be associated with the generator of symmetry. By expanding expression (\ref{eq:e8}) it is easy to verify that the conservation laws $Y^{(k)}$ can be expressed in terms of the integrals of motion $c_{1} ... c_{n}$ in the following way 
\begin{eqnarray} \label{eq:e9}
Y^{(k)} = \frac{ (n - k)! k! }{n! } \sum _{i_{p} \neq i_{s}} c_{i_{1}}c_{i_{2}} ... c_{i_{k}}
\end{eqnarray}
\\
{\bf Sample.} Let $M$ be $R^{4}$ with coordinates $z_{1}, z_{2}, z_{3}, z_{4}$ and Poisson bivector field 
\begin{eqnarray} \label{eq:e10}
W = \partial _{z_{1}} \wedge \partial _{z_{3}} + \partial _{z_{2}} \wedge \partial _{z_{4}}
\end{eqnarray}
($\partial _{z_{a}}$ just denotes derivative with respect to $z_{a}$ coordinate) and let's take 
\begin{eqnarray}
h = \frac{1}{2}z_{1}^{2} + \frac{1}{2}z_{2}^{2} + e^{z_{3} - z_{4}} 
\end{eqnarray}
Then the vector field 
\begin{eqnarray}
E = \sum ^{ 4 }_{a = 1 } E_{a} \partial _{z_{a}} 
\end{eqnarray}
with components 
\begin{eqnarray} \label{eq:e11}
E_{1} = \frac{1}{2}z_{1}^{2} - e^{z_{3} - z_{4}} - \frac{t}{2}(z_{1} + z_{2})e^{z_{3} - z_{4}}\nonumber \\E_{2} = \frac{1}{2}z_{2}^{2} + 2e^{z_{3} - z_{4}} + \frac{t}{2}(z_{1} + z_{2})e^{z_{3} - z_{4}}\nonumber \\E_{3} = 2z_{1} + \frac{1}{2}z_{2} + \frac{t}{2}(z_{1}^{2} + e^{z_{3} - z_{4}})\nonumber \\E_{4} = z_{2} - \frac{1}{2}z_{1} + \frac{t}{2}(z_{2}^{2} + e^{z_{3} - z_{4}}) 
\end{eqnarray}
satisfies (\ref{eq:e5}) condition and as a result generates symmetry of the dynamical system. The symmetry appears to be non-Noether with Schouten bracket $[E , W]$ equal to 
\begin{eqnarray} \label{eq:e12}
\hat{W} = [E , W] = z_{1}\partial _{z_{1}} \wedge \partial _{z_{3}} + z_{2}\partial _{z_{2}} \wedge \partial _{z_{4}} + e^{z_{3} - z_{4}}\partial _{z_{1}} \wedge \partial _{z_{2}} + \partial _{z_{3}} \wedge \partial _{z_{4}} 
\end{eqnarray}
calculating volume vector fields $\hat{W}^{k} \wedge W^{n - k}$ gives rise to 
\begin{eqnarray}
W \wedge W = - 2\partial _{z_{1}} \wedge \partial _{z_{2}} \wedge \partial _{z_{3}} \wedge \partial _{z_{4}}\nonumber \\\hat{W} \wedge W = - (z_{1} + z_{2})\partial _{z_{1}} \wedge \partial _{z_{2}} \wedge \partial _{z_{3}} \wedge \partial _{z_{4}}\nonumber \\\hat{W} \wedge \hat{W} = - 2(z_{1}z_{2} - e^{z_{3} - z_{4}}) \partial _{z_{1}} \wedge \partial _{z_{2}} \wedge \partial _{z_{3}} \wedge \partial _{z_{4}} 
\end{eqnarray}
and the conservation laws associated with this symmetry are just 
\begin{eqnarray}
Y^{(1)} = \frac{ \hat{W} \wedge W }{W \wedge W } = \frac{1}{2}(z_{1} + z_{2})\nonumber \\Y^{(2)} = \frac{ \hat{W} \wedge \hat{W} }{W \wedge W } = z_{1}z_{2} - e^{z_{3} - z_{4}}
\end{eqnarray}
\\
Presence of the non-Noether symmetry not only leads to a sequence of conservation laws, but also endows the phase space with a number of interesting geometric structures and it appears that such a symmetry is related to many important concepts used in theory of dynamical systems. One of the such concepts is Lax pair. Let us recall that Lax pair of Hamiltonian system on Poisson manifold $M$ is a pair $(L , P)$ of smooth functions on $M$ with values in some Lie algebra $g$ such that the time evolution of $L$ is governed by the following equation 
\begin{eqnarray} \label{eq:e13}
\frac{d}{dt}L = [L , P] 
\end{eqnarray}
where $[ , ]$ is a Lie bracket on $g$. It is well known that each Lax pair leads to a number of conservation laws. When $g$ is some matrix Lie algebra the conservation laws are just traces of powers of $L$ 
\begin{eqnarray} \label{eq:e14}
I^{(k)} = Tr(L^{k}) 
\end{eqnarray}
It is remarkable that each generator of the non-Noether symmetry canonically leads to the Lax pair of a certain type. In the local coordinates $z_{a}$, where the bivector field $W$ and the generator of the symmetry $E$ have the following form 
\begin{eqnarray}
W = \sum _{ab} W_{ab} \partial _{z_{a}} \wedge \partial _{z_{b}} ~~~~~ E = \sum _{a} E_{a} \partial _{z_{a}} 
\end{eqnarray}
corresponding Lax pair could be calculated explicitly. Namely we have the following theorem: \\
{\bf Theorem 2.} Let $(M , h)$ be regular Hamiltonian system on the $2n$-dimensional Poisson manifold $M$. Then, if the vector field $E$ on $M$ generates the non-Noether symmetry, the following $2n\times 2n$ matrix valued functions on $M$ 
\begin{eqnarray} \label{eq:e15}
L_{ab} = \sum _{dc} W^{- 1}_{ad}(E_{c}\partial _{z_{c}}W_{db} - W_{cb} \partial _{z_{c}}E_{d} + W_{dc} \partial _{z_{c}}E_{b})\nonumber \\P_{ab} = \sum _{c} \partial _{z_{a}}(W_{bc}\partial _{z_{c}}h) 
\end{eqnarray}
form the Lax pair (\ref{eq:e13}) of the dynamical system $(M , h)$. \\
{\bf Proof:} Let us consider the following operator on a space of 1-forms 
\begin{eqnarray} \label{eq:e16}
\bar R_{E}(u) = \Phi _{W}^{- 1}([E , \Phi _{W}(u)]) - L_{E}u 
\end{eqnarray}
(here $\Phi _{W}$ is the isomorphism (\ref{eq:e2})). It is obvious that $\bar R_{E}$ is a linear operator and it is invariant since time evolution commutes with both $\Phi _{W}$ (as far as $[W(h) , W] = 0$) and $E$ (because $E$ generates symmetry). In the terms of the local coordinates $\bar R_{E}$ has the following form 
\begin{eqnarray}
\bar R_{E} = \sum _{ab} L_{ab} dz_{a} \otimes \partial _{z_{b}} 
\end{eqnarray}
and the invariance condition 
\begin{eqnarray}
\frac{d}{dt}\bar R_{E} = L_{W(h)}\bar R_{E} = 0 
\end{eqnarray}
yields 
\begin{eqnarray}
\frac{d}{dt}\bar R_{E} = \frac{d}{dt}\sum _{ab} L_{ab} dz_{a} \otimes \partial _{z_{b}} = \sum _{ab} (\frac{d}{dt}L_{ab}) dz_{a} \otimes \partial _{z_{b}}\nonumber \\+ \sum _{ab} L_{ab} (L_{W(h)}dz_{a}) \otimes \partial _{z_{b}} + \sum _{ab} L_{ab} dz_{a} \otimes (L_{W(h)}\partial _{z_{b}} ) =\nonumber \\\sum _{ab} (\frac{d}{dt}L_{ab}) dz_{a} \otimes \partial _{z_{b}} + \sum _{abcd} L_{ab}\partial _{z_{c}}(W_{ad}\partial _{z_{d}}h)dz_{c} \otimes \partial _{z_{b}} + \sum _{abcd} L_{ab}\partial _{z_{b}}(W_{cd}\partial _{z_{d}}h)dz_{a} \otimes \partial _{z_{c}} =\nonumber \\\sum _{ab} (\frac{d}{dt}L_{ab} + \sum _{c} (P_{ac}L_{cb} - L_{ac}P_{cb})) dz_{a} \otimes \partial _{z_{b}} = 0 
\end{eqnarray}
or in matrix notations 
\begin{eqnarray}
\frac{d}{dt}L = [L , P]. 
\end{eqnarray}
So, we have proved that the non-Noether symmetry canonically yields a Lax pair on the algebra of linear operators on cotangent bundle over the phase space. \\
{\bf Remark.} The conservation laws (\ref{eq:e14}) associated with the Lax pair (\ref{eq:e13}) can be expressed in terms of the integrals of motion $c_{i}$ in quite simple way: 
\begin{eqnarray} \label{eq:e17}
I^{(k)} = Tr(L^{k}) = \sum _{i} c_{i}^{k}
\end{eqnarray}
This correspondence follows from the equation (\ref{eq:e8}) and the definition of the operator $\bar R_{E}$ (\ref{eq:e16}). \\
{\bf Sample.} Let us calculate Lax matrix associated with non-Noether symmetry (\ref{eq:e11}). Using (\ref{eq:e15}) it is easy to check that Lax matrix has eight nonzero elements 
\begin{eqnarray}
L_{11} = L_{33} = z_{1}\nonumber \\L_{22} = L_{44} = z_{2}\nonumber \\L_{32} = - L_{41} = e^{z_{3} - z_{4}}\nonumber \\L_{23} = - L_{14} = 1 
\end{eqnarray}
The conservation laws associated with this Lax matrix are 
\begin{eqnarray} \label{eq:e18}
I^{(1)} = Tr(L) = 2(z_{1} + z_{2})\nonumber \\I^{(2)} = Tr(L^{2}) = 2z_{1}^{2} + 2z_{2}^{2} + 4e^{z_{3} - z_{4}} 
\end{eqnarray}
\\
Now let us focus on the integrability issues. We know that $n$ integrals of motion are associated with each generator of non-Noether symmetry and according to the Liouville-Arnold theorem Hamiltonian system is completely integrable if it possesses $n$ functionally independent integrals of motion in involution (two functions $f$ and $g$ are said to be in involution if their Poisson bracket vanishes $\{f , g\} = 0$). Generally speaking the conservation laws associated with symmetry might appear to be neither independent nor involutive. However it is reasonable to ask the question -- what condition should be satisfied by the generator of the symmetry to ensure the involutivity ($\{Y^{(k)} , Y^{(m)}\} = 0$) of conserved quantities? In Lax theory such a condition is known as Classical Yang-Baxter Equation (CYBE). Since involutivity of the conservation laws is closely related to the integrability it is essential to have some analog of CYBE for the generator of non-Noether symmetry. To address this issue we would like to propose the following theorem. \\
{\bf Theorem 3.} If the vector field $E$ on $2n$-dimensional Poisson manifold $M$satisfies the condition 
\begin{eqnarray} \label{eq:e19}
[[E[E , W]]W] = 0
\end{eqnarray}
and $W$ bivector field has maximal rank ($W^{n} \neq 0$) then the functions (\ref{eq:e6}) are in involution 
\begin{eqnarray}
\{Y^{(k)} , Y^{(m)}\} = 0
\end{eqnarray}
\\
{\bf Proof:} First of all let us note that the identity (\ref{eq:e1}) satisfied by the Poisson bivector field $W$ is responsible for the Liouville theorem 
\begin{eqnarray} \label{eq:e20}
[W , W] = 0 ~~~~~\Leftrightarrow ~~~~~ L_{W(f)}W = [W(f) , W] = 0
\end{eqnarray}
By taking the Lie derivative of the expression (\ref{eq:e1}) we obtain another useful identity 
\begin{eqnarray}
L_{E}[W , W] = [E[W , W]] = [[E , W] W] + [W[E , W]] = 2[\hat{W} , W] = 0.
\end{eqnarray}
This identity gives rise to the following relation 
\begin{eqnarray} \label{eq:e21}
[\hat{W} , W] = 0 ~~~~~\Leftrightarrow ~~~~~ [\hat{W}(f) , W] = - [\hat{W} , W(f)]
\end{eqnarray}
and finally condition (\ref{eq:e19}) ensures third identity 
\begin{eqnarray}
[\hat{W} , \hat{W}] = 0
\end{eqnarray}
yielding Liouville theorem for $\hat{W}$ 
\begin{eqnarray} \label{eq:e22}
[\hat{W} , \hat{W}] = 0 ~~~~~\Leftrightarrow ~~~~~ [\hat{W}(f) , \hat{W}] = 0
\end{eqnarray}
Indeed 
\begin{eqnarray}
[\hat{W} , \hat{W}] = [[E , W]\hat{W}] = [[\hat{W} , E]W] = - [[E , \hat{W}]W] = - [[E[E , W]]W] = 0
\end{eqnarray}
Now let us consider two different solutions $c_{i} \neq c_{j}$ of the equation (\ref{eq:e8}). By taking the Lie derivative of the equation 
\begin{eqnarray}
(\hat{W} - c_{i}W)^{n} = 0
\end{eqnarray}
along the vector fields $W(c_{j})$ and $\hat{W}(c_{j})$ and using Liouville theorem for $W$ and $\hat{W}$ bivectors we obtain the following relations 
\begin{eqnarray} \label{eq:e23}
(\hat{W} - c_{i}W)^{n - 1}(L_{W(c_{j})}\hat{W} - \{c_{j} , c_{i}\}W) = 0,
\end{eqnarray}
and 
\begin{eqnarray} \label{eq:e24}
(\hat{W} - c_{i}W)^{n - 1}(c_{i}L_{\hat{W}(c_{j})}W + \{c_{j} , c_{i}\}_{\bullet }W) = 0,
\end{eqnarray}
where 
\begin{eqnarray}
\{c_{i} , c_{j}\}_{\bullet } = \hat{W}(dc_{i} \wedge dc_{j})
\end{eqnarray}
is the Poisson bracket calculated by means of the bivector field $\hat{W}$. Now multiplying (\ref{eq:e23}) by $c_{i}$ subtracting (\ref{eq:e24}) and using identity (\ref{eq:e21}) gives rise to 
\begin{eqnarray} \label{eq:e25}
(\{c_{i} , c_{j}\}_{\bullet } - c_{i}\{c_{i} , c_{j}\})(\hat{W} - c_{i}W)^{n - 1}W = 0
\end{eqnarray}
Thus, either 
\begin{eqnarray} \label{eq:e26}
\{c_{i} , c_{j}\}_{\bullet } - c_{i}\{c_{i} , c_{j}\} = 0
\end{eqnarray}
or the volume field $(\hat{W} - c_{i}W)^{n - 1}W$ vanishes. In the second case we can repeat (\ref{eq:e23})-(\ref{eq:e25}) procedure for the volume field $(\hat{W} - c_{i}W)^{n - 1}W$ yielding after $n$ iterations $W^{n} = 0$ that according to our assumption (that the dynamical system is regular) is not true. As a result we arrived at (\ref{eq:e26}) and by the simple interchange of indices $i \leftrightarrow j$ we get 
\begin{eqnarray} \label{eq:e27}
\{c_{i} , c_{j}\}_{\bullet } - c_{j}\{c_{i} , c_{j}\} = 0
\end{eqnarray}
Finally by comparing (\ref{eq:e26}) and (\ref{eq:e27}) we obtain that the functions $c_{i}$ are in involution with respect to the both Poisson structures (since $c_{i} \neq c_{j}$) 
\begin{eqnarray}
\{c_{i} , c_{j}\}_{\bullet } = \{c_{i} , c_{j}\} = 0
\end{eqnarray}
and according to (\ref{eq:e9}) the same is true for the integrals of motion $Y^{(k)}$. \\
{\bf Corollary.} Each generator of non-Noether symmetry satisfying equation (\ref{eq:e19}) endows dynamical system with the bi-Hamiltonian structure -- couple ($W , \hat{W}$) of compatible ($[W , \hat{W}] = 0$) Poisson ($[W , W] = [\hat{W} , \hat{W}] = 0$) bivector fields. \\
{\bf Remark.} Theorem 3 is useful in multidimentional dynamical systems where involutivity of conservation laws can not be checked directly.\\
{\bf Sample.} One can check that the non-Noether symmetry (\ref{eq:e11}) satisfies condition (\ref{eq:e19}) and the bivector fields $W$ and $\hat{W}$ defined by (\ref{eq:e10}) and (\ref{eq:e12}) form bi-Hamiltonian system $[W , W] = [W , \hat{W}] = [\hat{W} , \hat{W}] = 0$. \\
Another concept that is often used in theory of dynamical systems and could be related to the non-Noether symmetry is the bidifferential calculus (bicomplex approach). Recently A.~Dimakis and F.~M\"{u}ller-Hoissen applied bidifferential calculi to the wide range of integrable models including KdV hierarchy, KP equation, self-dual Yang-Mills equation, Sine-Gordon equation, Toda models, non-linear Schr\"{o}dinger and Liouville equations. It turns out that these models can be effectively described and analyzed using the bidifferential calculi \cite{r1} \cite{r2}. \\
Under the bidifferential calculus we mean the graded algebra of differential forms 
\begin{eqnarray}
\Omega = \cup ^{ \infty }_{k = 0 } \Omega ^{(k)} 
\end{eqnarray}
($\Omega ^{(k)}$ denotes the space of $k$-degree differential forms) equipped with a couple of differential operators 
\begin{eqnarray}
d, \tilde{d} : \Omega ^{(k)} \rightarrow \Omega ^{(k + 1)} 
\end{eqnarray}
satisfying $d^{2} = \tilde{d}^{2} = d\tilde{d} + \tilde{d}d = 0$ conditions (see \cite{r2}). It is interesting that if generator of the non-Noether symmetry satisfies equation (\ref{eq:e19}) then we are able to construct an invariant bidifferential calculus of a certain type. This construction is summarized in the following theorem: \\
{\bf Theorem 4.} Let $(M , h)$ be regular Hamiltonian system on the Poisson manifold $M$. Then, if the vector field $E$ on $M$ generates the non-Noether symmetry and satisfies the equation (\ref{eq:e19}), the differential operators 
\begin{eqnarray} \label{eq:e28}
du = \Phi _{W}^{- 1}([W , \Phi _{W}(u)])
\end{eqnarray}
\begin{eqnarray} \label{eq:e29}
\tilde{d}u = \Phi _{W}^{- 1}([[E , W]\Phi _{W}(u)])
\end{eqnarray}
form invariant bidifferential calculus ($d^{2} = \tilde{d}^{2} = d\tilde{d} + \tilde{d}d = 0$) over the graded algebra of differential forms on $M$. \\
{\bf Proof:} First of all we have to show that $d$ and $\tilde{d}$ are really differential operators , i.e., they are linear maps from $\Omega ^{(k)}$ into $\Omega ^{(k + 1)}$, satisfy derivation property and are nilpotent ($d^{2} = \tilde{d}^{2} = 0$). Linearity is obvious and follows from the linearity of the Schouten bracket $[ , ]$ and $\Phi _{W}, \Phi _{W}^{- 1}$ maps. Then, if $u$ is a $k$-degree form $\Phi _{W}$ maps it on $k$-degree multivector field and the Schouten brackets $[W , \Phi _{W}(u)]$ and $[[E , W]\Phi _{W}(u)]$ result the $k + 1$-degree multivector fields that are mapped on $k + 1$-degree differential forms by $\Phi _{W}^{- 1}$. So, $d$ and $\tilde{d}$ are linear maps from $\Omega ^{(k)}$ into $\Omega ^{(k + 1)}$. Derivation property follows from the same feature of the Schouten bracket $[ , ]$ and linearity of $\Phi _{W}$ and $\Phi _{W}^{- 1}$ maps. Now we have to prove the nilpotency of $d$ and $\tilde{d}$. Let us consider $d^{2}u$ 
\begin{eqnarray}
d^{2}u = \Phi _{W}^{- 1}([W , \Phi _{W}(\Phi _{W}^{- 1}([W , \Phi _{W}(u)]))]) = \Phi _{W}^{- 1}([W[W , \Phi _{W}(u)]]) = 0
\end{eqnarray}
as a result of the property (\ref{eq:e20}) and the Jacobi identity for $[ , ]$ bracket. In the same manner 
\begin{eqnarray}
\tilde{d}^{2}u = \Phi _{W}^{- 1}([[W , E][[W , E]\Phi _{W}(u)]]) = 0
\end{eqnarray}
according to the property (\ref{eq:e22}) of $[W , E] = \hat{W}$ and the Jacobi identity. Thus, we have proved that $d$ and $\tilde{d}$ are differential operators (in fact $d$ is ordinary exterior differential and the expression (\ref{eq:e28}) is its well known representation in terms of Poisson bivector field). It remains to show that the compatibility condition $d\tilde{d} + \tilde{d}d = 0$ is fulfilled. Using definitions of $d, \tilde{d}$ and the Jacobi identity we get 
\begin{eqnarray}
(d\tilde{d} + \tilde{d}d)(u) = \Phi _{W}^{- 1}([[[W , E]W]\Phi _{W}(u)]) = 0 
\end{eqnarray}
as far as (\ref{eq:e21}) is satisfied. So, $d$ and $\tilde{d}$ form the bidifferential calculus over the graded algebra of differential forms. It is also clear that the bidifferential calculus $d, \tilde{d}$ is invariant, since both $d$ and $\tilde{d}$ commute with time evolution operator $W(h) = \{h, \}$. \\
{\bf Remark.} Conservation laws that are associated with the bidifferential calculus (\ref{eq:e28}) (\ref{eq:e29}) and form Lenard scheme (see \cite{r2}): 
\begin{eqnarray}
(k + 1)\tilde{d}I^{(k)} = kdI^{(k + 1)}
\end{eqnarray}
coincide with the sequence of integrals of motion (\ref{eq:e17}). Proof of this correspondence lay outside the scope of present article, but could be done in the manner similar to \cite{r1}. \\
{\bf Sample.} The symmetry (\ref{eq:e11}) endows $R^{4}$ with bicomplex structure $d, \tilde{d}$ where $d$ is ordinary exterier derivative while $\tilde{d}$ is defined by 
\begin{eqnarray}
\tilde{d}z_{1} = z_{1}dz_{1} - e^{z_{3} - z_{4}}dz_{4}\nonumber \\\tilde{d}z_{2} = z_{2}dz_{2} + e^{z_{3} - z_{4}}dz_{3}\nonumber \\\tilde{d}z_{3} = z_{1}dz_{3} + dz_{2}\nonumber \\\tilde{d}z_{4} = z_{2}dz_{4} - dz_{1} 
\end{eqnarray}
and is extended to whole De Rham complex by linearity, derivation property and compatibility property $d\tilde{d} + \tilde{d}d = 0$. The conservation laws $I^{(1)}$ and $I^{(2)}$ defined by (\ref{eq:e18}) form the simpliest Lenard scheme 
\begin{eqnarray}
2\tilde{d}I^{(1)} = dI^{(2)}
\end{eqnarray}
\\
Finally we would like to reveal some features of the operator $\bar R_{E}$ (\ref{eq:e16}) and to show how Fr\"{o}licher-Nijenhuis geometry could arise in Hamiltonian system that possesses certain non-Noether symmetry. From the geometric properties of the tangent valued forms we know that the traces of powers of a linear operator $F$ on tangent bundle are in involution whenever its Fr\"{o}licher-Nijenhuis torsion $T(F)$ vanishes, i. e. whenever for arbitrary vector fields $X,Y$ the condition 
\begin{eqnarray}
T(F)(X , Y) = [FX , FY] - F([FX , Y] + [X , FY] - F[X , Y]) = 0
\end{eqnarray}
is satisfied. Torsionless forms are also called Fr\"{o}licher-Nijenhuis operators and are widely used in theory of integrable models. We would like to show that each generator of non-Noether symmetry satisfying equation (\ref{eq:e19}) canonnically leads to invariant Fr\"{o}licher-Nijenhuis operator on tangent bundle over the phase space. Strictly speaking we have the following theorem. \\
{\bf Theorem 5.} Let $(M , h)$ be regular Hamiltonian system on the Poisson manifold $M$. If the vector field $E$ on $M$ generates the non-Noether symmetry and satisfies the equation (\ref{eq:e19}) then the linear operator, defined for every vector field $X$ by equation 
\begin{eqnarray}
R_{E}(X) = \Phi _{W}(L_{E}\Phi _{W}^{- 1}(X)) - [E , X] 
\end{eqnarray}
is invariant Fr\"{o}licher-Nijenhuis operator on $M$. \\
{\bf Proof.} Invariance of $R_{E}$ follows from the invariance of the $\bar R_{E}$ defined by (\ref{eq:e16}) (note that for arbitrary 1-form vector field $u$ and vector field $X$ contraction $i_{X}u$ has the property $i_{R_{E}X}u = i_{X}\bar R_{E}u$, so $R_{E}$ is actually transposed to $\bar R_{E}$). It remains to show that the condition (\ref{eq:e19}) ensures vanishing of the Fr\"{o}licher-Nijenhuis torsion $T(R_{E})$ of $R_{E}$, i.e. for arbitrary vector fields $X, Y$ we must get 
\begin{eqnarray} \label{eq:e30}
T(R_{E})(X , Y) = [R_{E}(X) , R_{E}(Y)] - R_{E}([R_{E}(X) , Y] + [X , R_{E}(Y)] - R_{E}([X , Y])) = 0 
\end{eqnarray}
First let us introduce the following auxiliary 2-forms 
\begin{eqnarray} \label{eq:e31}
\omega = \Phi _{W}^{- 1}(W), ~~~~~ \omega ^{\bullet } = \bar R_{E}\omega ~~~~~ \omega ^{\bullet \bullet } = \bar R_{E}\omega ^{\bullet }
\end{eqnarray}
Using the realization (\ref{eq:e28}) of the differential $d$ and the property (\ref{eq:e1}) yields 
\begin{eqnarray}
d\omega = \Phi _{W}^{- 1}([W , W]) = 0
\end{eqnarray}
Similarly, using the property (\ref{eq:e21}) we obtain 
\begin{eqnarray}
d\omega ^{\bullet } = d\Phi _{W}^{- 1}([E , W]) - dL_{E}\omega = \Phi _{W}^{- 1}([[E , W]W]) - L_{E}d\omega = 0
\end{eqnarray}
And finally, taking into account that $\omega ^{\bullet } = 2\Phi _{W}^{- 1}([E , W])$ and using the condition (\ref{eq:e19}), we get 
\begin{eqnarray}
d\omega ^{\bullet \bullet } = 2\Phi _{W}^{- 1}([[E[E , W]]W]) - 2dL_{E}\omega ^{\bullet } = - 2L_{E}d\omega ^{\bullet } = 0
\end{eqnarray}
So the differential forms $\omega , \omega ^{\bullet }, \omega ^{\bullet \bullet }$ are closed 
\begin{eqnarray} \label{eq:e32}
d\omega = d\omega ^{\bullet } = d\omega ^{\bullet \bullet } = 0
\end{eqnarray}
Now let us consider the contraction of $T(R_{E})$ and $\omega $. 
\begin{eqnarray} \label{eq:e33}
i_{T(R_{E})(X , Y)}\omega = i_{[R_{E}X , R_{E}Y]}\omega - i_{[R_{E}X , Y]}\omega ^{\bullet } - i_{[X , R_{E}Y]}\omega ^{\bullet } + i_{[X , Y]}\omega ^{\bullet \bullet } = \nonumber \\L_{R_{E}X}i_{Y}\omega ^{\bullet } - i_{R_{E}Y}L_{X}\omega ^{\bullet } - L_{R_{E}X}i_{Y}\omega ^{\bullet } + i_{Y}L_{R_{E}X}\omega ^{\bullet } - L_{X}i_{R_{E}Y}\omega ^{\bullet } + i_{R_{E}Y}L_{X}\omega ^{\bullet } + i_{[X , Y]}\omega ^{\bullet \bullet } = \nonumber \\i_{Y}L_{X}\omega ^{\bullet \bullet } - L_{X}i_{Y}\omega ^{\bullet \bullet } + i_{[X , Y]}\omega ^{\bullet \bullet } = 0 
\end{eqnarray}
where we used (\ref{eq:e31}) (\ref{eq:e32}), the property of the Lie derivative 
\begin{eqnarray}
L_{X}i_{Y}\omega = i_{Y}L_{X}\omega + i_{[X , Y]}\omega 
\end{eqnarray}
and the relations of the following type 
\begin{eqnarray}
L_{R_{E}X}\omega = di_{R_{E}X}\omega + i_{R_{E}X}d\omega = di_{X}\omega ^{\bullet } = L_{X}\omega ^{\bullet } - i_{X}d\omega ^{\bullet } = L_{X}\omega ^{\bullet } 
\end{eqnarray}
So we proved that for arbitrary vector fields $X, Y$ the contraction of $T(R_{E})(X , Y)$ and $\omega $ vanishes. But since $W$ bivector is non-degenerate ($W^{n} \neq 0$), its counter image 
\begin{eqnarray}
\omega = \Phi _{W}^{- 1}(W)
\end{eqnarray}
is also non-degenerate and vanishing of the contraction (\ref{eq:e33}) implies that the torsion $T(R_{E})$ itself is zero. So we get 
\begin{eqnarray}
T(R_{E})(X , Y) = [R_{E}(X) , R_{E}(Y)] - R_{E}([R_{E}(X) , Y] + [X , R_{E}(Y)] - R_{E}([X , Y])) = 0 
\end{eqnarray}
\\
{\bf Sample.} Note that operator $R_{E}$ associated with non-Noether symmetry (\ref{eq:e11}) reproduces well known Fr\"{o}licher-Nijenhuis operator 
\begin{eqnarray}
R_{E} = z_{1}dz_{1} \otimes \partial _{z_{1}} - dz_{1} \otimes \partial _{z_{4}} + z_{2}dz_{2} \otimes \partial _{z_{2}} + dz_{2} \otimes \partial _{z_{3}} + \nonumber \\z_{1}dz_{3} \otimes \partial _{z_{3}} + e^{z_{3} - z_{4}}dz_{3} \otimes \partial _{z_{2}} + z_{2}dz_{4} \otimes \partial _{z_{4}} - e^{z_{3} - z_{4}}dz_{4} \otimes \partial _{z_{1}} 
\end{eqnarray}
(compare with \cite{r3}) \\
In summary let us note that the non-Noether symmetries form quite interesting class of symmetries of Hamiltonian dynamical system and lead not only to a number of conservation laws (that under certain conditions ensure integrability), but also enrich the geometry of the phase space by endowing it with several important structures, such as Lax pair, bicomplex, bi-Hamiltonian structure, Fr\"{o}licher-Nijenhuis operators etc. The present paper attempts to emphasize deep relationship between different concepts used in construction of conservation laws and non-Noether symmetry. \\
\section{Acknowledgements}
Author is grateful to Zakaria Giunashvili, George Jorjadze and Michael Maziashvili for constructive discussions and help. This work was supported by INTAS (00-00561).\\

\end{document}